\documentclass[preprint]{elsarticle}
\usepackage{natbib}
\usepackage{lineno}
\usepackage{amsmath}
\usepackage{indentfirst}
\usepackage{amsfonts}
\usepackage{amssymb}
\usepackage{changepage}
\usepackage{lscape}
\usepackage[]{algorithm2e}
\usepackage{afterpage}
\usepackage[T1]{fontenc}
\usepackage{breqn}
\usepackage{booktabs}
\usepackage{hyperref}

\usepackage{subfigure}
\usepackage{color}
\usepackage{epsfig}
\usepackage{epstopdf}
\usepackage{amsmath}
\usepackage{graphicx,psfrag,epsf}
\usepackage{afterpage}
\usepackage{color}
\usepackage{soul}
\usepackage{multirow}
\usepackage{rotating}
\usepackage{scrextend}
\usepackage{comment}

\newcommand{\bTheta}{\boldsymbol\Theta}
\newcommand{\btheta}{\boldsymbol\theta}

\newcommand{\bE}{\mathbf{E}}

\newcommand{\bX}{\mathbf{X}}

\newcommand{\tT}{\mathrm{T}}

\newcommand{\Expect}{{\rm I\kern-.3em E}}

\newdefinition{defin}{Definition}

\begin{document}
\begin{frontmatter}
\title{\textbf{Population Power Curves in ASCA with Permutation Testing}}

\author{Jos{\'e} Camacho\corref{cor1}\fnref{fn1}}
\author{Michael Sorochan Armstrong\fnref{fn1}}

\ead{josecamacho@ugr.es} 
\fntext[fn1]{Research Centre for Information and Communication Technologies (CITIC-UGR), University of Granada, Spain}
\cortext[cor1]{Corresponding author: josecamacho@ugr.es} 

\date{}

\begin{abstract}
In this paper, we revisit the Power Curves in ANOVA Simultaneous Component Analysis (ASCA) based on permutation testing, and introduce the Population Curves derived from population parameters describing the relative effect among factors and interactions. { The relative effect has important practical implications: the statistical power of a given factor depends on the design of other factors in the experiment, and not only of the sample size. Thus, understanding the relative power in a specific experimental design can be extremely useful to maximize our capability of success when planning the experiment. In the paper, we derive Relative and Absolute Population Curves}, where the former represent statistical power in terms of the normalized effect size between structure and noise, and the latter in terms of the sample size.  {Both types of population curves allow us to make decisions regarding the number and nature (fixed/random) of factors, their relationships (crossed/nested), the number of levels and replicates, among others, in an multivariate experimental design (e.g., an omics study) during the planning phase of the experiment.} We illustrate both types of curves through simulation. 

\end{abstract} 
\end{frontmatter}

\section{Introduction}

Prof. Smilde's research group proposed the ANOVA-Simultaneous Component Analysis (ASCA) \cite{smilde2005anova}, a powerful framework for analyzing the individual influence of different experimental factors and their interactions in experimental designs with a high number of responses. ASCA represents a natural multivariate extension of the Analysis of Variance (ANOVA). The theory associated to ANOVA is extensive \cite{montgomery2020design}, and several of its developments have been incrementally incorporated into the ASCA framework throughout the years \cite{thiel2017asca+,martin2020limm,madssen2021repeated}. Still, ASCA can be considered a developing technique, and many unresolved questions remain regarding best practices \cite{camacho2023permutation,polushkina}.

One such questions is how to derive a power analysis in the context of ASCA. Statistical power is a relevant concept in inferential statistics. The power of a test measures the probability that it correctly rejects the null hypothesis ($H_{0}$) when the alternative hypothesis ($H_{1}$) is true. The power can be defined as $1-\beta$ for $\beta$ the Type II Error (false negative) probability. Power Curves are a form of power analysis where power is represented in terms of the sample size, that is the number of replicates or experimental runs under the same conditions. This is a recommended analysis prior to any multivariate experiment (e.g., a clinical study with omics responses) to determine the required number of subjects and experimental levels for a desired statistical power.

In standard univariate ANOVA, or when relatively few responses are considered, it is possible to use analytical methods to derive Power Curves based on typical assumptions (such as normality) \cite{practicalstats2018}. However, in the context of more than just a few responses, or to assess the violation of any possible statistical assumptions, numerical methods \cite{anderson2003permutation} are a viable alternative. Arguably the  most popular approach for statistical inference in ASCA is permutation testing \cite{anderson2003permutation,vis2007statistical}: a  resampling method that transforms ASCA into a distribution-free approach that is more flexible than parametric ANOVA, {where inference is based on analytical distributions}. The derivation of Power Curves based on ASCA's permutation testing necessarily requires numerical approaches.

In previous work \cite{camacho2023permutation}, we introduced the simulated Power Curves in ASCA as a strategy to optimize an ASCA pipeline for a specific experimental design in terms of statistical power. Simulated Power Curves can be used to find the optimal ASCA model in terms of fixed/random factors, crossed/nested relationships, interactions, test statistic, transformations, and others. Our approach was defined to compare several models in terms of relative power, but it cannot be used to make sample size estimations for an entirely new multivariate experiment. In this paper, we revisit the simulated Power Curve approach to generalize it. In particular, we make the following contributions:
\begin{itemize}
    \item We generalize ASCA Power Curves so that any design can be simulated, including complex relationships among factors and their interactions. 
    
    \item We define the Population Curves, derived from population parameters (standard deviations) describing the relative effect among factors and interactions.
    \item We propose two types of Population Curves:
    \begin{itemize}
    \item Relative Population Curves represent statistical power in terms of the relative effect size between structure and noise. They are useful to optimize the ASCA pipeline for an analysis at hand.
    \item Absolute Population Curves represent statistical power in terms of the sample size. They are useful to plan ahead the number of replicates and/or levels to use in a designed study.
    \end{itemize}
    \item We illustrate the behaviour of the two types of Population Curves through simulation. 
    \item We provide open software for the generation of Population Curves and for the replication of the results in this paper. Relative and Absolute Population Curves can be computed with `powercurve' routine in the MEDA Toolbox stable release v1.4\footnote{Stable release of MEDA Toolbox v1.4. \url{https://github.com/CoDaSLab/MEDA\-Toolbox/releases/tag/v1.4}}. 
\end{itemize}

The rest of the paper is organized as follows: Section \ref{sec:ASCA} introduces the ASCA framework. Section  \ref{sec:PC} discusses the concept of Power Curves from a theoretical perspective.  Sections \ref{sec:RPC} and  \ref{sec:APC} present the algorithms to compute Relative and Absolute Population Curves, respectively. Section \ref{sec:Examples} discusses simulation results and Section \ref{sec:Conclusion} draws the conclusions of the work.

\section{ANOVA Simultaneous Component Analysis} \label{sec:ASCA}

ASCA, like ANOVA, is mostly concerned with the analysis of data coming from an experimental design. Following ANOVA, a common ASCA pipeline follows three steps: (1) factorization of the data according to the experimental design; (2) significance testing for factors and interactions; (3) visualization of significant {factors' and interactions' effects} using Principal Component Analysis (PCA) to understand separability among levels.

\subsection{Factorization of the data}

Let $\bX$ be {an} $N\times M$ data matrix with $N$ {the number of experimental runs} and $M$ {the number of} responses in a designed experiment. Without loss of generality, we consider here the case of a design with two crossed factors $A$ and $B$, their interaction, and an additional factor $C(A)$ nested in $A$. This is a common configuration in multiple omics experiments~\cite{koleini2023complementary,martin2020limm,madssen2021repeated}, useful to correct for the (often large) individual variability ($C(A)$) and so increase the statistical power of the test. The data in $\bX$ can be decomposed as: 
{\begin{equation}
	\label{eqn:RepMes}
	\bX = \mathbf{1}\mathbf{m}^{\tT} + \mathbf{X}_{A} + \mathbf{X}_{B} + \mathbf{X}_{C(A)} +\mathbf{X}_{AB} + \bE
\end{equation}
\noindent where $\mathbf{1}$ is a vector of ones {($N\times 1$)}, $\mathbf{m}$ {($M\times 1$) denotes a vector containing the intercepts of the {$M$} measured variables}, $\mathbf{X}_A$, $\mathbf{X}_{B}$ and $\mathbf{X}_{C(A)}$ represent the factor matrices, $\mathbf{X}_{AB}$} is the interaction matrix, and $\bE$ is the residual matrix, all of similar dimensions of $\bX$.

To compute the factorization, we use the technique referred to as ASCA+ \citep{thiel2017asca+} to account for mild unbalancedness in the data. We intentionally avoid ASCA alternatives based on Linear Mixed Models (LMM) \cite{martin2020limm,madssen2021repeated} due to the large increase of computational demand, prohibitive in the context of Power Curves. In ASCA+, the decomposition is derived as the least squares solution of a regression problem, where $\bX$ is regressed onto a coding matrix $\mathbf{D}$ {as:} 
\begin{equation}
\label{eqn:asca+}
\bX = \mathbf{D}\bTheta  + \bE =  \mathbf{1}\btheta + \mathbf{D}_A\bTheta_A + \mathbf{D}_B\bTheta_B + \mathbf{D}_{C(A)}\bTheta_{C(A)} + \mathbf{D}_{AB}\bTheta_{AB} + \bE
\end{equation}
\noindent and {$\mathbf{D}$} is constructed using sum coding \citep{thiel2017asca+} or another alternative coding schemes \citep{madssen2021repeated} and $\bTheta$ and $\bE$ are obtained from:
{
\begin{equation}
\label{eqn:ls}
\bTheta =  (\mathbf{D}^{\tT}\mathbf{D})^{-1}\mathbf{D}\mathbf{X}
\end{equation}
\begin{equation}
\label{eqn:ls2}
\bE =  \mathbf{X} - \mathbf{D}\bTheta 
\end{equation}}
The encoding of experimental factors in the design matrix is an especially important consideration, as it affects what information is passed through to variance explained by the model, versus residual variance--this in turn affects the apparent evidence for statistical significance.

\subsection{Statistical significance testing}

{In this step, we test the statistical significance of factors and interactions in a similar way as performed in multi-way ANOVA.} A widely used approach for ASCA inference is permutation testing \cite{anderson2003permutation,vis2007statistical}. Permutation testing can be performed by randomly shuffling the rows of $\mathbf{X}$ in Equation (\ref{eqn:ls}), yielding a new set of regression coefficients, and so a new factorization:
\begin{equation}
\label{eq:perraw}
\bTheta^* =  (\mathbf{D}^{\tT}\mathbf{D})^{-1}\mathbf{D}^{\tT}\mathbf{X}^*
\end{equation}
\begin{equation}
\bE^* =  \mathbf{X}^* - \mathbf{D}\bTheta^* 
\end{equation}
Permutation tests are carried out by comparing a given statistic, computed {after the ASCA factorization}, 
with the corresponding statistic computed from hundreds or more permutations of the rows in the observational data, $\mathbf{X}$. The $p$-value is obtained as\footnote{Throughout the paper, we assume that the higher the statistic the more significant {the effect of the factor/interaction.}}
\begin{equation}
\label{eq:$p$-value}
p =  \frac{\#\{S_k^* \geq S; k=1,\ldots,K\} + 1}{K+1}
\end{equation}
%
\noindent where $S$ refers to the statistic computed from the true factorization, $S_k^*$ is the statistic corresponding to the $k$-th random permutation, {$\#\{cond\}$ refers to the number of times condition $cond$ {is met},} and $K$ is the total number of permutations. See \cite{camacho2023permutation} for a recent review on the permutation approach and the relevance of the chosen statistic.

{The permutation approach in the rows of $\mathbf{X}$ allow to test all the factors and interactions at the same time. Alternatively, one may be interested in testing an individual factor/interaction, which can be done by permuting the corresponding coding rather than the data \cite{camacho2023permutation}, or by orthogonalizing the data prior to permutation. Additional considerations may be taken into account for imbalanced designs \cite{Morten}.}

\subsection{Post-hoc visualization}

{Significance testing in ANOVA/ASCA reveals the statistical significance of factors' and interactions' effects, but not of the specific levels (or combination thereof). To identify significant differences across levels, post-hoc tests are typically employed. The equivalent to a post-hoc visualization in ASCA is the use of PCA following Zwanenburg \emph{et al.}~\citep{zwanenburg2011anova}, where the PCA loadings are computed from the factor/interaction matrix, and score plots are built from the sum of this matrix and the residuals. Following this approach, score plots provide a visual comparison between the main/interaction effects and the natural variability in the residuals. Other approaches besides PCA can also be used \cite{ARMSTRONG202461}.

ASCA is a supervised method, and as such it can suffer of ``overfitting'' like any other regression and/or classification approaches \cite{babyak2004you}. ASCA relies in permutation testing to avoid over-fitting. Thus, only statistically significant factors/interactions should be visualized post-hoc with PCA.

\section{Power Curves} \label{sec:PC}

Figure \ref{fig:xplan} illustrates the concept of a power curve as a comparison between two distributions: the null distribution $P(t|H_0)$ and the alternative distribution $P(t|H_1)$. In the figure, both distributions are assumed to be normal, but in a real situation this {may not} be the case. The shaded areas represent the probability of error, associated with an observation incorrectly rejecting the null hypothesis through random chance (type I error), with probability $\alpha$ (in blue), versus the probability of an observation incorrectly not rejecting the null hypothesis (type II error), with probability $\beta$ (in red). Larger effect sizes $\theta$, manifesting as wider separations between both distributions, increase the statistical power $1-\beta$ (reducing the type II error) for a fixed value of $\alpha$.  Alternatively, for both fixed $\alpha$ and the effect size $\theta$, increasing the sampling size reduces the variance of the distributions $P(t|H_0)$ and $P(t|H_1)$, which leads to a reduction of overlapping areas as $1-\beta$ increases.

\begin{figure}
    \centering
    \includegraphics{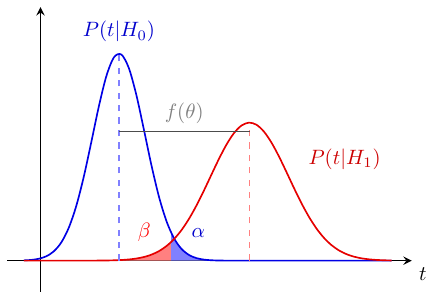}
    \caption{Illustration of effect size ($f(\theta)$) versus the probability of a type-I error, $\alpha$, and the probability of a type II error, $\beta$. Statistical power is defined as 1 - the false negative (type-II) error rate.}
    \label{fig:xplan}
\end{figure}

\section{Relative Population Curves for ASCA} \label{sec:RPC}

Relative Population Curves or RPCs are derived from population parameters describing the standard deviation in factors and interactions and represent statistical power in terms of the relative effect size between structure and noise. 

To generate RPCs we generalize the approach by Camacho  \emph{et al.} \cite{camacho2023permutation}. We simulate data that progressively
increases the relative effect size $\theta$ and therefore the power (Figure \ref{fig:xplan}). In the following, we use model (\ref{eqn:RepMes}) to showcase the RPCs, since this model includes crossed relationships, interactions (between $A$ and $B$) and nested relationships (between $A$ and $C(A)$), so that almost any other model can be derived from it. We can also think of this model as an illustration of factors/interactions organized in different orders (Figure \ref{fig:Hasse}), as discussed by Anderson and Ter Braak \cite{anderson2003permutation} and in the Hasse diagrams by Marini \emph{et al.} \cite{marini2015analysis}.

\begin{figure*}
 	\centering
        \includegraphics[width=0.6\textwidth]{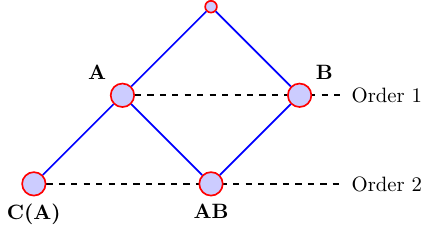}
        
 	\caption{Ordering structure among factors and interaction in model (\ref{eqn:RepMes}).}
\label{fig:Hasse}
\end{figure*}

Our approach to generate RPCs follows these steps:

{
\begin{itemize}
	\item [0.] INPUT:

	\begin{itemize}
		\item $\mathbf{F}$ the design matrix
            \item $H$ a model hierarchical structure (as illustrated by a Hassel diagram) with information about factors, crossed/nested relationships and their interactions. In this model structure, a factor $f$ can have descendants (nested factors and associated interactions) and ancestors (factors in which $f$ is nested). An interaction $i$ can also have descendants (other higher-order interactions) and ancestors ({other lower-order interactions or} the factors included in the interaction). In model (\ref{eqn:RepMes}) and Figure \ref{fig:Hasse}, $A$ is ancestor of $C(A)$ and $A$ and $B$ are ancestors of $AB$.
            \item $M$ the number of responses
            \item $k_f$ for $f \in \{1 ... F\}$, $k_i$ for $i \in \{1 ... I\}$, and $k_e$ the coefficients with the standard deviation\footnote{Given the multivariate nature of the response, and to simplify notation, we generally refer to the standard deviation $\sigma$ of each response vector, so that the expected standard deviation of each individual response would be $\sigma/\sqrt{M}$} for factors and the interactions, and the residuals
		\item $R$ the number of repetitions to generate a statistically representative RPC
		\item $P$ the number of permutations (in each repetition)
	    \item $\delta$ incremental steps in the effect size
		\item $\alpha$ the imposed probability of falsely rejecting the null hypothesis (i.e. the significance level)
\end{itemize}

	\item [1.] Set $power_f(\theta) = 0$ and  $power_i(\theta) = 0$ for each factor $f$ and interaction $i$, respectively, and for $\theta$ from 0 to $10\delta$ in $\delta$ steps; and $N$ is set to the number of rows of $\mathbf{F}$.
	
	\item [2.] For each repetition from 1 to $R$
	
	\begin{itemize}
	
	\item [2.1.] Generate random matrices to represent level/cell averages

	\begin{itemize}
 
        \item[2.1.1.] For each factor $f_1$ of order 1 in $H$, and so with no ancestors (e.g., $A$ and $B$ in model (\ref{eqn:RepMes})), count the number of levels $L_{f_1}$ in matrix $\mathbf{F}$ and simulate:
        \begin{equation}
        \mathbf{\bar{X}}_{f_1}(L_{f_1},M) \sim \mathcal{D}_{f_1}
        \end{equation}
        where $\mathcal{D}_{f_1}$ is a pseudo-random number generator (PRNG), potentially based on the normal distribution or other distribution that may deviate moderately (e.g., uniform) or severely (e.g., exp$^3$) from normality~\cite{anderson2003permutation}. 
    
	\item[2.1.2.] For each factor $f$  in $H$ with ancestor factor(s) $f_a$, for $a = {1...A_f}$ (e.g., $C(A)$ nested on $A$ in model (\ref{eqn:RepMes})):
        \begin{equation}
         L_{f} = r_f \cdot \prod_{a=1}^{A_f} L_{f_a}
        \end{equation}
        \begin{equation}
        \label{eq:barfac}
        \mathbf{\bar{X}}_{f}(L_{f},M) \sim \mathcal{D}_{f}
        \end{equation}
        with $r_f$ the number of replicates in each unique combination of levels (cell), and $\mathcal{D}_{f}$ the chosen PRNG.
        
	\item[2.1.3.] For each interaction $i$ in $H$ with ancestor factor(s) $f_a$ for $a = {1...A_i}$  (e.g., $AB$ in model (\ref{eqn:RepMes})):
        \begin{equation}
         L_{i} = \prod_{a=1}^{A_i} L_{i_a}
        \end{equation}
	\begin{equation}
        \label{eq:barint}
	\mathbf{\bar{X}}_{i}(L_{i},M) \sim \mathcal{D}_i
	\end{equation}
        with $\mathcal{D}_{i}$ the chosen PRNG.

	\end{itemize}
        
	\item [2.2.] Generate background variability for a chosen PRNG:
	\begin{equation}
        \label{eq:res}
	\mathbf{X}_E(N,M) \sim \mathcal{D}_E
	\end{equation}
        \item [2.3.] Normalize each matrix $\mathbf{\bar{X}}_f$ for each factor $f$, $\mathbf{\bar{X}}_i$ for each interaction $i$, and $\mathbf{X}_E$ so that the Frobenius norm equals the squared root of the number of rows\footnote{This normalization is instrumental for the correspondence of theoretical and numerical results, as shown later on.}.
        
	\item [2.4.] For each observation $n$ from 1 to $N$: 

 	\begin{itemize}
 
        \item[2.4.1.]  For each factor $f$, we build $\mathbf{X}_{f}(N,M)$ from $\mathbf{\bar{X}}_{f}(L_f,M)$ and the design matrix $\mathbf{F}$:
		\begin{equation}
		\mathbf{x}_f^n = \mathbf{\bar{x}}^{l}_f 
		\end{equation}
        with $\mathbf{x}_f^n$ the $n$-th row of $\mathbf{X}_{f}(N,M)$ and $\mathbf{\bar{x}}^{l}_f$ the $l$-th row of $\mathbf{\bar{X}}_{f}(L_f,M)$, with $l$ determined according to matrix $\mathbf{F}$.

        \item[2.4.2.] For each interaction $i$, we build $\mathbf{X}_{i}(N,M)$ from $\mathbf{\bar{X}}_{i}(L_i,M)$ and the design matrix $\mathbf{F}$:
		\begin{equation}
		\mathbf{x}_i^n = \mathbf{\bar{x}}^{l}_i 
		\end{equation}
        with $\mathbf{x}_i^n$ the $n$-th row of $\mathbf{X}_{i}(N,M)$ and $\mathbf{\bar{x}}^{l}_i$ the $l$-th row of $\mathbf{\bar{X}}_{i}(L_i,M)$, with $l$ determined according to matrix $\mathbf{F}$.
        
	\end{itemize}
        
	\item [2.5.] Compute the matrices with the structural and residual part with the standard deviation coefficients:
	\begin{equation}
        \label{eq:str}
	\mathbf{X}_S  =  \sum_f k_f \mathbf{X}_f + \sum_i k_i \mathbf{X}_i 
	\end{equation}
	\begin{equation}
        \label{eq:noi}
	\mathbf{X}_E  =  k_e \mathbf{X}_E 
	\end{equation}
	\item [2.6.] For $\theta$ from 0 to $10\delta$ in $\delta$ steps:
	\begin{itemize}
		\item [2.6.1.] Yield the simulated data:
		{\begin{equation}
            \label{eq:final}
		\mathbf{X} = \theta \mathbf{X}_{S} + \mathbf{X}_E
		\end{equation}}
		\item [2.6.2.] Compute ASCA+ partition and the F-ratio for factors and interactions for both the simulated data and $P$ permutations. For (high-order) factors and interactions in $H$ with no descendants:
		\begin{equation}
        \label{eq:f}
		F_f = (SS_{f} / DoF_f) / (SS_{E} / DoF_E) 
		\end{equation}
		\begin{equation}
		F_i = (SS_{i} / DoF_i) / (SS_{E} / DoF_E) 
		\end{equation}
            where $SS$ refers to the sum-of-squares (the Frobenius norm) of a factor/interaction/residual matrix in the factorization with ASCA+, and $DoFs$ represents the corresponding degrees of freedom \cite{montgomery2020design}. The DoFs of a factor is the number of levels minus one, the DoFs of an interaction is the product of the DoFs of its factors, and the DoFs of the residuals is the total (number of observations minus 1 in the data) minus the DoFs of all factors and interactions in the model. The ratio between the $SS$ and the $DoF$ is often called the mean sum-of-squares ($MS$).
            For any factors and interactions in $H$ with descendants:
		\begin{equation}
        \label{eq:approx}
		F_f = \frac{(SS_{f} / DoF_f)}{(\sum_{d=1}^{D_f} SS_{f_d} + \sum_{d=1}^{D_i} SS_{i_d})/(\sum_{d=1}^{D_f} DoF_{f_d} + \sum_{d=1}^{D_i} DoF_{i_d})} 
		\end{equation}
		\begin{equation}
  \label{eq:fend}
		F_i = \frac{(SS_{i} / DoF_i)}{(\sum_{d=1}^{D_i} SS_{i_d})/(\sum_{d=1}^{D_i} DoF_{i_d})) } 
		\end{equation}
            with descendant factor(s) $f_d$ for $d = {1...D_f}$ and descendant interaction(s) $f_i$ for $d = {1...D_i}$.
            
		\item [2.6.3.] For each factor $f$, if the associated p-value computed using Eq. (\ref{eq:$p$-value}) is below {$\alpha$}  do:
		\begin{equation}
		      power_f(\theta) = power_f(\theta) + 1
		\end{equation}
		\item [2.6.4.] For each factor $i$, if the associated p-value computed using Eq. (\ref{eq:$p$-value}) is below {$\alpha$}  do:
		\begin{equation}
		      power_i(\theta) = power_i(\theta) + 1
		\end{equation}
	\end{itemize}
	\end{itemize} 

	\item [3.] Normalize $power_f(\theta) = power_f(\theta)/R$ and  $power_i(\theta) = power_i(\theta)/R$ for each factor $f$ / interaction $i$

\end{itemize} }

The algorithm works as follows. In step 0, we set the general characteristics of the data simulation. In step 1, the algorithm initializes the values in the RPC. In step 2, we iterate through a number of repetitions to compute the RPC. Each repetition consists on the simulation of a structural part (in $\mathbf{X}_{S}$) and a residual part (in $\mathbf{X}_{E}$). The inner loop builds the data for intermediate cases between the absence of effect ($\mathbf{X} = \mathbf{X}_{E}$) and absence of residuals ($\mathbf{X} = \mathbf{X}_{S}$), factorizes it with ASCA+ and performs the statistical inference through permutation testing. If the computed significance is below the imposed significance level, the power is increased {by} one. In step 3, the power is normalized by the number of repetitions.

Some alternative configurations for a RPC that are relevant in practice can be straightforwardly implemented by modifying specific parameters or small parts of the algorithm:

\begin{itemize}

\item Non-balanced designs can be easily integrated in the design matrix $\mathbf{F}$, see ~\cite{camacho2023permutation}.

\item By properly choosing $\mathcal{D}_E$, we can emulate different distributions in the residuals to assess robustness {to} deviations from normality in the manner of Anderson and Ter Braak~\cite{anderson2003permutation} but for multivariate responses. We can also generate the level averages using different distributions.

\item Both fixed and random factors can be simulated in the same manner.

\item We can integrate complex designs by adding multiple crossed and nested relationships as well as interactions.

\item We can generate RPCs for a specific factor or interaction in an experimental design, or for all of them. The algorithm provides the solution for a RPC that considers a simultaneous incremental effect in all factors and interactions of the model, but some factors may be deactivated by setting the corresponding {standard deviation} coefficient to 0. We can also maintain the effect of a significant factor/interaction fixed along the curve by adding its contribution directly in Eq.~(\ref{eq:final}), for example: 
{\begin{equation}
		\mathbf{X} = \theta \mathbf{X}_{S} +  \mathbf{X}_E + k_f \mathbf{X}_f
		\end{equation}}
\item In a similar way, we can add a covariate $\mathbf{X}_{cv}(N,M) \sim \mathcal{D}_{cv}$ directly in Eq.~(\ref{eq:final}), for example: 
{\begin{equation}
		\mathbf{X} = \theta \mathbf{X}_{S} +  \mathbf{X}_E + k_{cv} \mathbf{X}_{cv}
		\end{equation}}
\item We can generate RPCs for alternative statistics {to the F-ratio in Eqs. (\ref{eq:f})-(\ref{eq:fend})}, see~\cite{camacho2023permutation}.

\end{itemize}

\section{Absolute Population Curves for ASCA} \label{sec:APC}

Absolute Population Curves (APCs) differ to RPCs in that the power is shown in terms of the sampling size, rather than the relative effect size. APCs have the same applications as the RPCs, but with the additional advantage that they provide information about the number of replicates and/or factor levels that one may use in a multivariate experiment in order to attain a given probability of success, that is, the probability of rejecting the null hypothesis when the alternative hypothesis is true: the power $1-\beta$. 

An APC is built by simulating data that progressively enlarges the number of levels in a specific factor, or the whole experimental design. The consequence of this enlargement is a reduction of the variance of both the null and the alternative distributions (Figure \ref{fig:xplan}), with a subsequent increase of statistical power. For instance, if we take model (\ref{eqn:RepMes}), we can apply APCs to investigate how the statistical power of the factors and the interaction is affected when:

\begin{itemize}
    \item We iteratively enlarge the number of levels in $A$. Often $A$ is a factor that controls the number of groups of individuals in a clinical study, e.g., with a number of disease sub-types. This APC gives useful information about whether incorporating more or less sub-types can impact our probability of success.
    
    \item We iteratively enlarge the number of levels in $B$. Often $B$ is a factor with several repeated measures over the same individual, e.g., in time or in biological samples. Then, the APC will allow us to investigate the effect of adding more time points/biological samples in our probability of success.
    
    \item We iteratively enlarge the number of levels in $C(A)$. Often $C(A)$ models the individual variability. The APC will give us an idea about the number of individuals per group we should choose for a certain probability of success.
    
    \item We iterative enlarge the whole experiment. The APC will give us an idea about the general number of replicates we should choose for a certain probability of success. This is often a good alternative choice when a factor like $C(A)$ is not in the design.
\end{itemize}

Our approach to generate APCs follows these steps (the algorithm is summarized not to replicate detailed explanations of the RPC algorithm) 

{
\begin{itemize}
	\item [0.] INPUT:

	\begin{itemize}
		\item INPUTs in the RPC algorithm
            \item $\theta$ the fixed effect size
            \item $f_{rep}$ the index of the factor that is replicated, or 0 if all the whole experiment is replicated
	    \item $\delta$ incremental steps in the sampling size $\eta$
        \end{itemize}

	\item [1.] Set $power_f(\eta) = 0$ and  $power_i(\eta) = 0$ for each factor $f$ and interaction $i$, respectively, and for $\eta$ from 1 to $10\delta$ in $\delta$ steps.
	
	\item [2.] For each repetition from 1 to $R$
	
	\begin{itemize}

 	\item [2.1.] For $\sigma$ from 1 to $10\delta$ in $\delta$ steps:
	   
        \begin{itemize}
		
        \item [2.1.1] Create $\mathbf{F}_{\theta}$ from $\mathbf{F}$ and $f_{rep}$; and $N$ is set to the number of rows of $\mathbf{F}$.
	
	\item [2.1.2] Generate random matrices to represent level averages in factors and interactions.
        
	\item [2.2.3] Generate background variability for a chosen PRNG.
 
        \item [2.2.4] Normalize each matrix $\mathbf{\bar{X}}_f$ for each factor $f$, $\mathbf{\bar{X}}_i$ for each interaction $i$, and $\mathbf{X}_E$ so that the Frobenius norm equals the squared root of the number of rows.
        
	\item [2.2.5] Compute final factor and interaction matrices of  $N$ rows.

	\item [2.2.6] Compute the matrices with the structural and residual part with the standard deviation coefficients.

	\item [2.2.7] Yield the simulated data.

	\item [2.2.8] Compute ASCA+ partition and the F-ratio for factors and interactions for both the simulated data and $P$ permutations.	
 
        \item [2.2.9] Update the power of factors and interactions.

	\end{itemize}
	\end{itemize} 

	\item [3.] Normalize $power_f(\eta) = power_f(\eta)/R$ and  $power_i(\eta) = power_i(\eta)/R$ for each factor $f$ / interaction $i$

\end{itemize} }

The algorithm works as follows: In step 0 we set the general characteristics of the data simulation, and in step 1 the algorithm initializes the values in the APC. In step 2 we iterate through a number of repetitions to compute the APC. Each repetition consists of 
an inner loop that progressively increases the sampling size by either increasing in one the number of levels of a given factor (for $f_{rep}>0$) or by adding one complete set of experimental runs. Subsequently, the same simulation approach as in an RPC is followed to generate the simulated matrix with both structural and residuals parts. This matrix is then factorized using ASCA+ and the statistical inference is performed through permutation testing. If the computed significance is below the imposed significance level, the power is increased in one. In step 3, the power is normalized by the number of repetitions.

\section{Simulation examples} \label{sec:Examples}

\subsection{Relative Population Curves}

\begin{figure*}
 	\centering
	\includegraphics[width=0.6\textwidth]{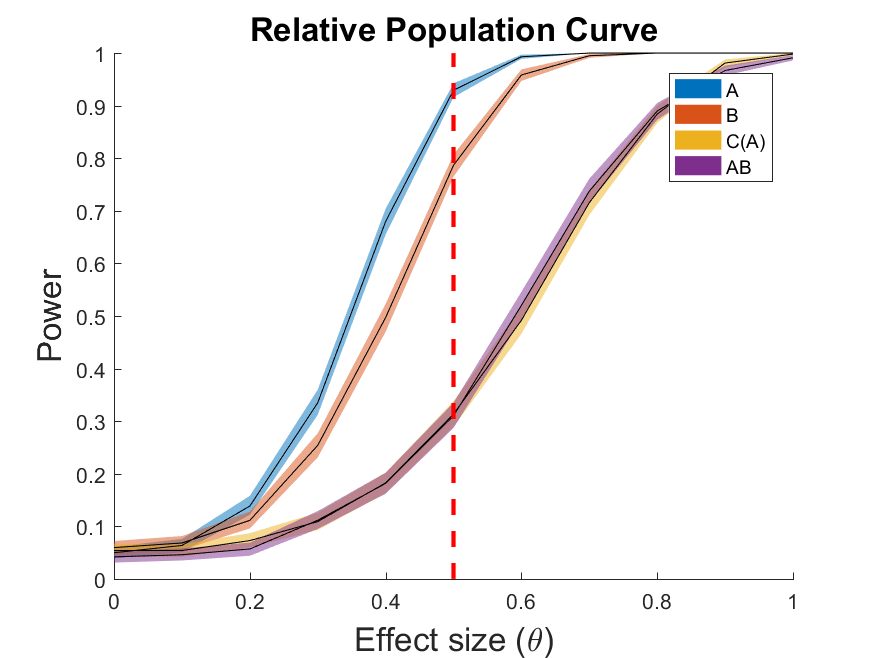}
 	\caption{Example of Relative Population Curve for model (\ref{eqn:RepMes}). The design matrix $\mathbf{F}$ contains a full factorial design with four levels in $A$, three levels in $B$ and four individuals in each cell of $C(A)$. Other inputs are $M=400$, $k_A = k_B = k_{C(A)} = k_{AB} = 0.2$, and $k_E = 1$, $R = 1000$, $P = 200$, $\delta=0.1$ and $\alpha = 0.05$. We marked $\theta = 0.5$ as a reference for the following figures. }
\label{fig:RPC1}
\end{figure*}

Figure \ref{fig:RPC1} presents an example of the RPCs for model (\ref{eqn:RepMes}) and for a full factorial design with four levels in $A$, three levels in $B$ and four individuals in each cell of $C(A)$. The RPCs are shown with $95\%$ confidence intervals computed by bootstrapping. In the example, all standard deviation coefficients are fixed to $1/5$ of the standard deviation in the residuals ({$k_A = k_B = k_{C(A)} = k_{AB} = 0.2$} and $k_E = 1$). The behaviour of the RPCs is the one expected for a correct power curve \cite{camacho2023permutation}: a) in the absence of effect (i.e., at $\theta=0$), all curves adjust to the significance level of $\alpha = 0.05$; and b) at some given effect size, the curves start gaining power until they reach 1.

{Interestingly, the order in which the curves in Figure \ref{fig:RPC1} 
 start rising and finally reach 1, that is, the relative power of the curves for the different factors and the interaction, is not intuitive given the same standard deviation of $0.2$ was used for all of them: $A$ is the most statistically powerful factor, followed by $B$, and $C(A)$ and $AB$ are the least powerful. We found that the relative power is a complex function of the ordering structure among factors/interactions (as depicted in Figure \ref{fig:Hasse}) and the number of levels thereof. The relative power has important practical implications, which are well-known in the area of design of experiments but not so widely understood by some experimenters: the statistical power of a given factor depends on the design of other factors in the experiment. To give an example, our ability to determine biological differences between a disease and a control group (factor $A$) depends on the number of individuals we include in the experiment (factor $C(A)$), but also on the number of repeated measures we take for each of them (factor $B$ and interaction $AB$). Thus, understanding the relative power in a specific experimental design can be extremely useful to maximize our capability of success. Generally speaking, the relative power is complicated to derive mathematically, especially in the presence of complex and varying null distributions across multiple responses, missing data, and other practicalities. Thus, RPCs are an interesting computational alternative for such derivation which can be made as specific to the problem at hand as desired. 

 While the mathematical derivation of multivariate power curves is often hopeless, we can still derive expected values for the variance in factors, interactions and residuals~\cite{montgomery2020design}, and so of the F-ratios. \ref{Math} provides such derivation for the example considered in Figure \ref{fig:RPC1}.} Using Eqs. (\ref{eq:approxA2})-(\ref{eq:AB2}), we can plot the expected F-ratios for the set of values of $\theta$, as illustrated in Figure \ref{fig:F}(a), and compare them to the averaged F-ratios obtained from the 1000 datasets simulated to compute the RPCs in Figure \ref{fig:RPC1}. These averaged F-ratios are in Figure \ref{fig:F}(b). We can see that the theoretical and numerical results match perfectly, which shows that our simulation approach accurately follows the ANOVA theory. We can also see that the F-ratios alone cannot explain the relative power observed in the RPCs of Figure \ref{fig:RPC1}: for instance, the F-ratio profiles of $A$ and $B$ are similar, while the relative power in the RPCs are not.  

\begin{figure*}
 	\centering
	\subfigure[Expected F-ratio]{\includegraphics[width=0.45\textwidth]{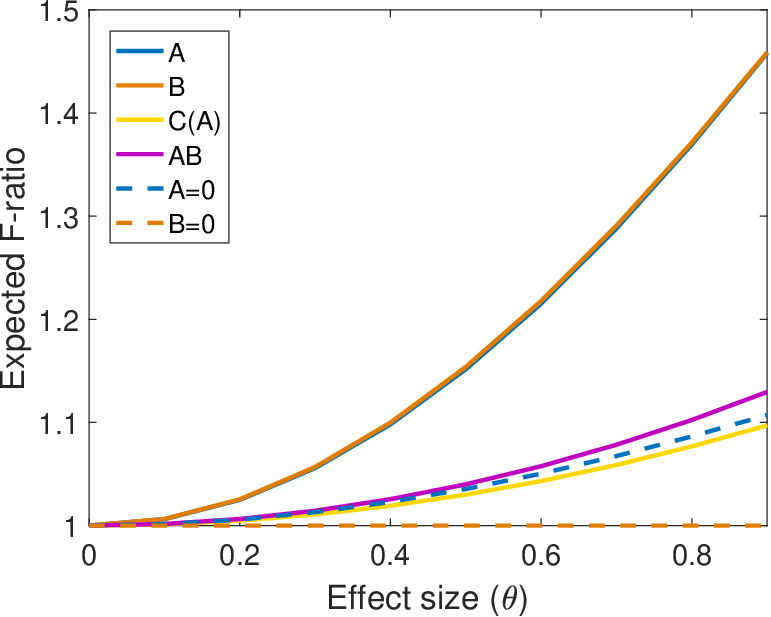}}
	\subfigure[Mean F-ratio]{\includegraphics[width=0.45\textwidth]{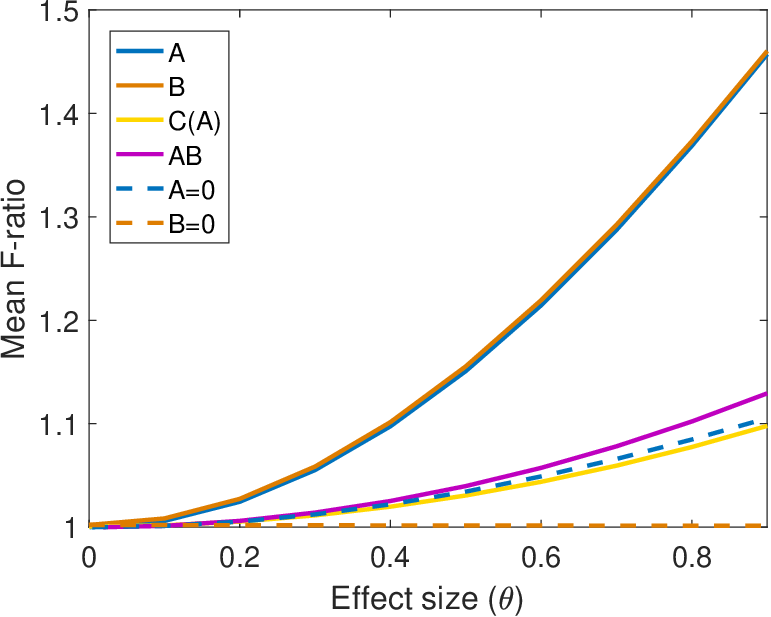}}

 	\caption{ Theoretical expected F-ratio (a) and mean simulated F-ratio (b) in terms of the effect size for the RPC in Figure \ref{fig:RPC1}.}
\label{fig:F}
\end{figure*}

\begin{figure*}
 	\centering
{\includegraphics[width=0.6\textwidth]{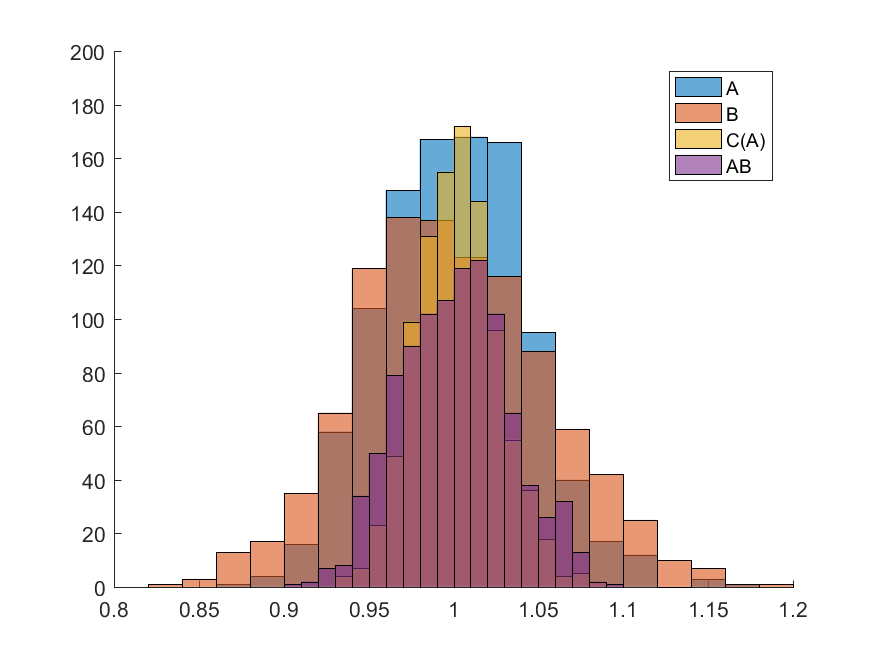}}

 	\caption{Null distributions for the first simulated dataset in the RPC in Figure \ref{fig:RPC1} and for $\theta = 0.5$.  }
\label{fig:Null}
\end{figure*}

The discrepancy between F-ratio and RPC profiles is caused by the different null distributions of factors and interactions. {Actually, it is the complexity to mathematically derive these null distributions which makes our computational approach a suitable tool to compute power curves. We illustrate the null distributions of our example in Figure \ref{fig:Null}}, generated with permutation testing for the first of the 1000 simulated datasets in the RPC and for $\theta = 0.5$. The null distribution of $B$ is significantly wider than the others. Since the p-value is obtained by comparing the F-ratio to the null distribution, and $A$ and $B$ show similar F-ratios at $\theta = 0.5$ (Figure \ref{fig:F}), the wider null distribution in $B$ makes the power curve to rise slower than that in $A$ in Figure \ref{fig:RPC1}. { This is because statistical power is associated to lower p-values. The ASCA table for the same simulated dataset in Figure \ref{fig:Null} is shown in Table \ref{table:ASCA}. The F-ratios and the p-values in the table are consistent with what we see in Figures \ref{fig:F} and \ref{fig:RPC1}, respectively, for $\theta = 0.5$: $A$ and $B$ present larger F-ratios that the others but still close to 1, $A$ is statistically significant while the rest are not. Note that this selected dataset in Figure \ref{fig:Null} and Table \ref{table:ASCA} is a single instance of the distribution that is averaged in Figures \ref{fig:RPC1} and \ref{fig:F}, which is the reason why some discrepancy is expected (e.g., $B$ is expected to be statistically significant 80\% of times at $\theta = 0.5$, according to Figure \ref{fig:RPC1}, but it is not in this example)}. 

{
\begin{table} 
\begin{tabular}{llllllll}
 & SumSq & PercSumSq & df & MeanSq & F & Pvalue \\ 
 \hline 
Mean & 1.3861 & 2.7832 & 1 & 1.3861 &  &  \\ 
A & 3.4739 & 6.9757 & 3 & 1.158 & 1.1096 & 0.00999 \\ 
B & 2.184 & 4.3855 & 2 & 1.092 & 1.0613 & 0.14585 \\ 
C(A) & 12.6117 & 25.3243 & 12 & 1.051 & 1.0522 & 0.023976 \\ 
AB & 6.1738 & 12.397 & 6 & 1.029 & 1.0302 & 0.17582 \\ 
Residuals & 23.9712 & 48.1342 & 24 & 0.9988 &  &  \\ 
Total & 49.8006 & 100 & 48 & 1.0375 &  &  \\ 
\end{tabular}
\caption{ASCA table for the first simulated dataset in the RPC of Figure \ref{fig:RPC1} and for $\theta = 0.5$.}
\label{table:ASCA}
\end{table} 

}

Figure \ref{fig:RPC2} presents several RPC examples similar to the first one, but where some of the factors or the interaction are deactivated with a null standard deviation. The figure shows that for any case where $k_B = 0$, $k_{C(A)} = 0$ and/or $k_{AB} = 0$, the corresponding RPC 
{stays at expected type I error of 0.05. This behaviour is not found for factor A  (so that the RPC does not go to 0.05 even for null effect in the factor) because its F-ratio is an approximate test rather than an exact one \cite{anderson2003permutation, montgomery2020design}\footnote{Please, note an exact test for $A$ in the experimental design of model (\ref{eqn:RepMes}) does not exist.}. These profiles in the RPCs can be explained from the theoretical derivation in \ref{Math}. Using this derivation, we included in Figure \ref{fig:F} the evolution of the expectation for the F-ratio of $A$ and $B$ for a null effect of the corresponding factor, marked with the labels $A=0$ and $B=0$, respectively. We can see that the approximate F-ratio in $A$ does not cancel out for a null variance of the factor, 
which leads} to the misleading RPC of $A$ that shows an unrealistic power. This undesirable behaviour in the {test of factor $A$, and so in the corresponding power curve, remains even when the variance of $C(A)$ or $AB$ is also cancelled out (Figures \ref{fig:RPC2}(f) and \ref{fig:RPC2}(g), respectively), and the power curves only works as expected when both are cancelled or when either $C(A)$ or $AB$ is not considered in the model, so that the F-ratio of $A$ corresponds to an exact test (Figure \ref{fig:RPC2}(h)). The RPC is a very useful tool to identify these situations, that is, when an approximate test provides unrealistic statistical power or lack of thereof, allowing us to avoid false negatives in practice: see \cite{camacho2023permutation,diaz2022predicting} for an example.}

\begin{figure*}
 	\centering
	\subfigure[$k_B = k_{C(A)} = k_{AB} = 0.2$, $k_A = 0$]{\includegraphics[width=0.45\textwidth]{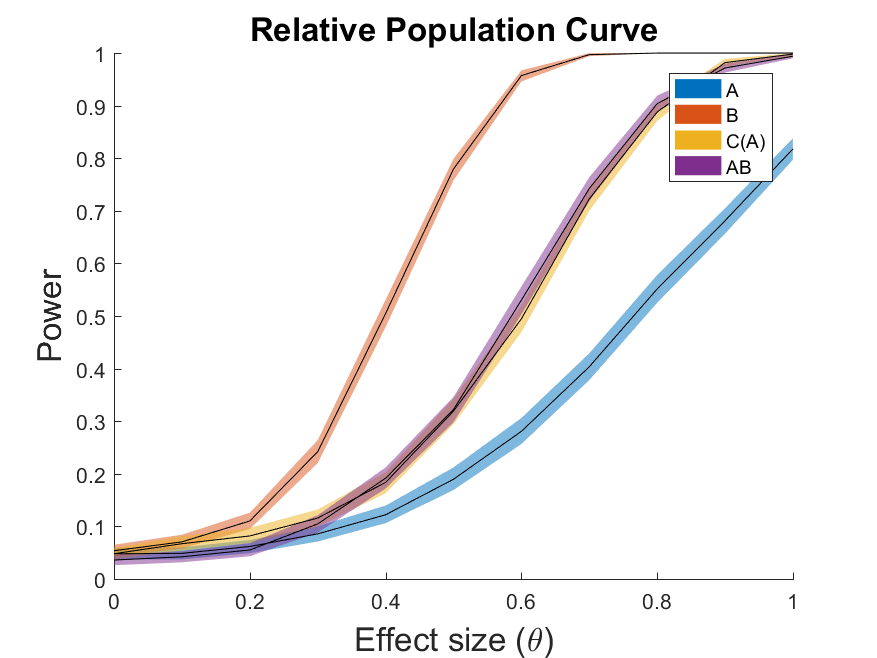}}
	\subfigure[$k_A = k_{C(A)} = k_{AB} = 0.2$, $k_B = 0$]{\includegraphics[width=0.45\textwidth]{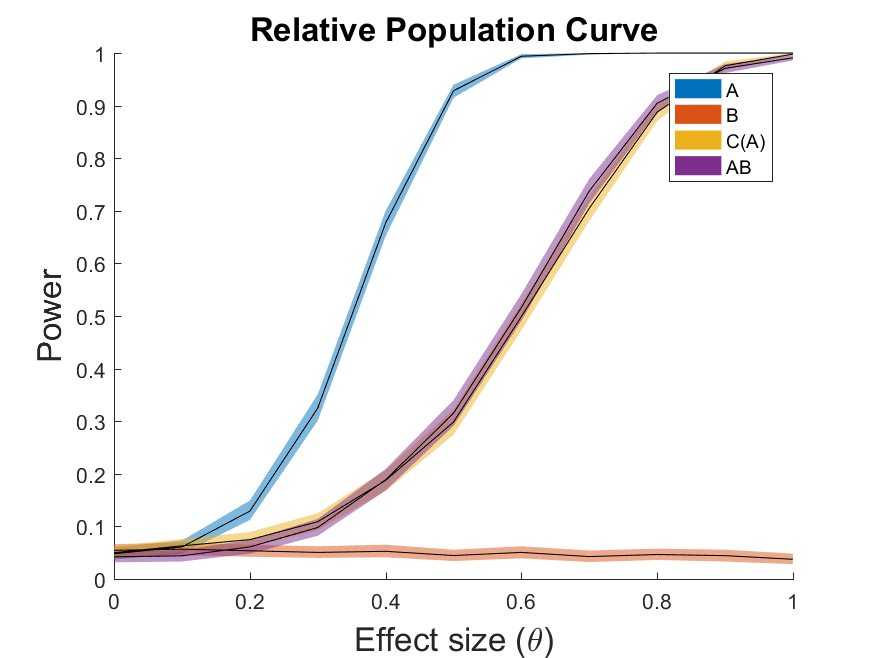}}
	\subfigure[$k_A = k_B = k_{AB} = 0.2$, $k_{C(A)} = 0$]{\includegraphics[width=0.45\textwidth]{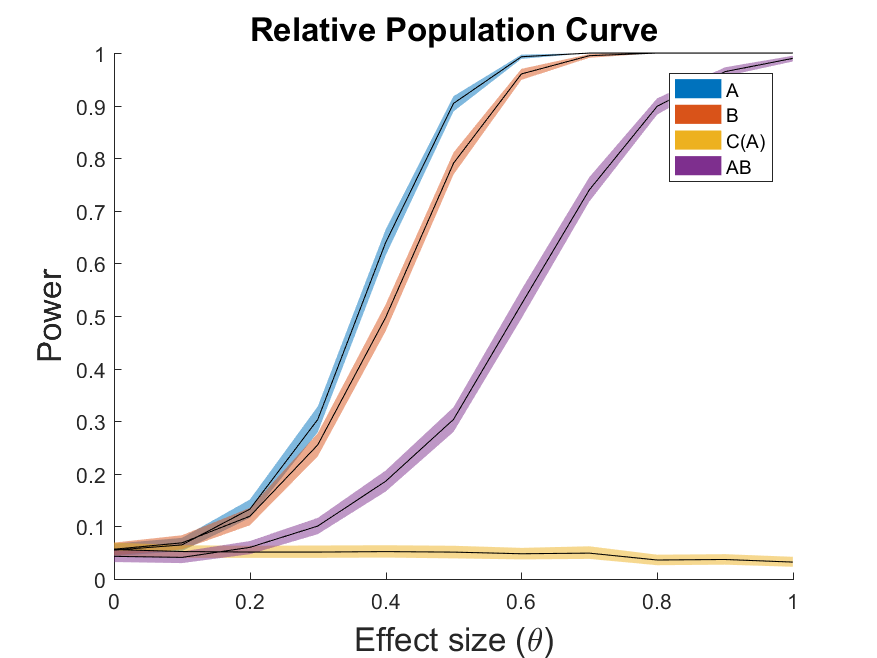}}
	\subfigure[$k_A = k_B = k_{C(A)} = 0.2$, $k_{AB} = 0$]{\includegraphics[width=0.45\textwidth]{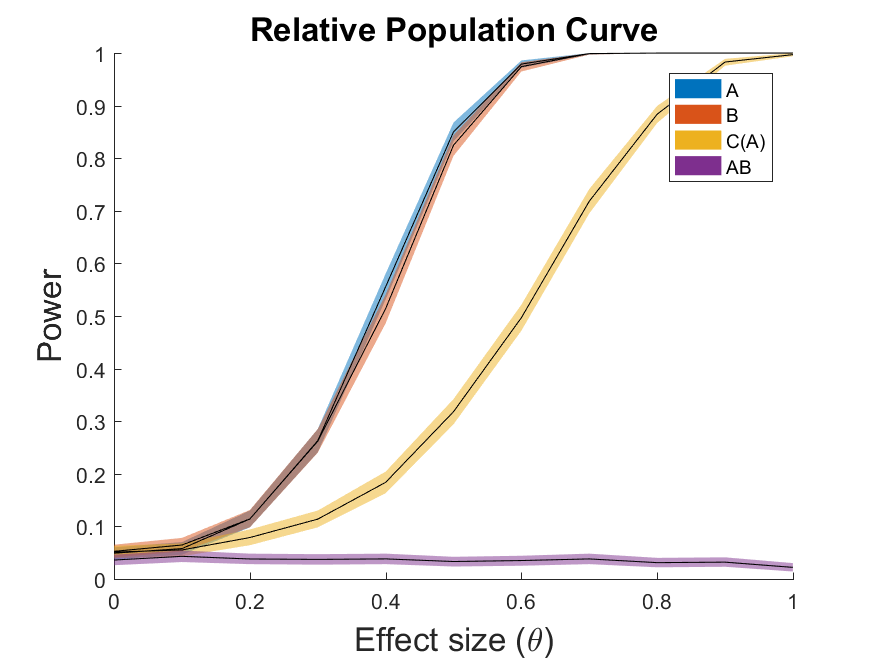}}
	\subfigure[$k_A = k_B =  0.2$, $k_{C(A)} = k_{AB} = 0$]{\includegraphics[width=0.45\textwidth]{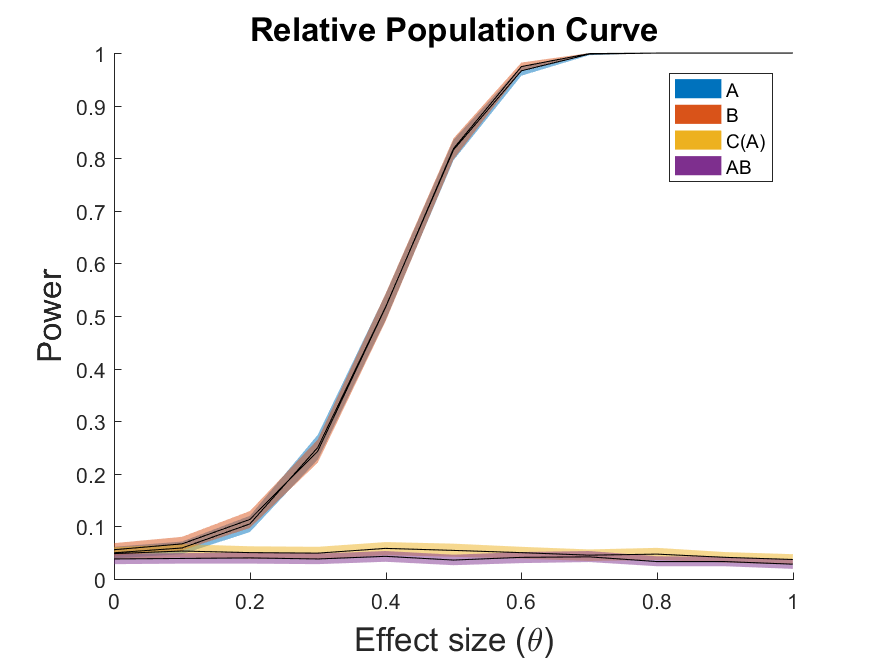}}
	\subfigure[$k_B = k_{AB} = 0.2$, $k_{A} = k_{C(A)} = 0$]{\includegraphics[width=0.45\textwidth]{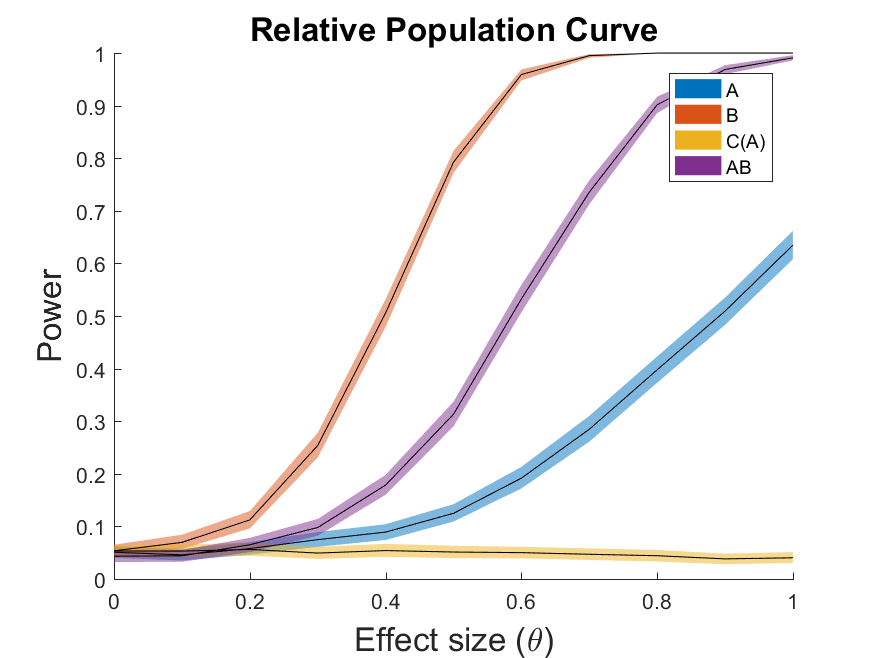}}
	\subfigure[$k_B = k_{C(A)} = 0.2$, $k_{A} = k_{AB} = 0$]{\includegraphics[width=0.45\textwidth]{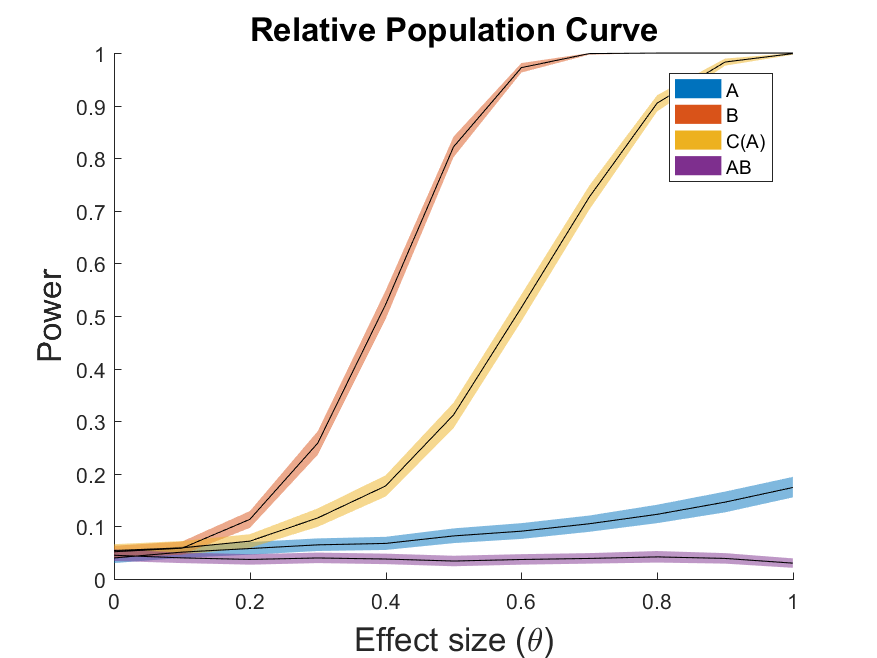}}
	\subfigure[$k_B = k_{AB} = 0.2$, $k_{A} = 0$]{\includegraphics[width=0.45\textwidth]{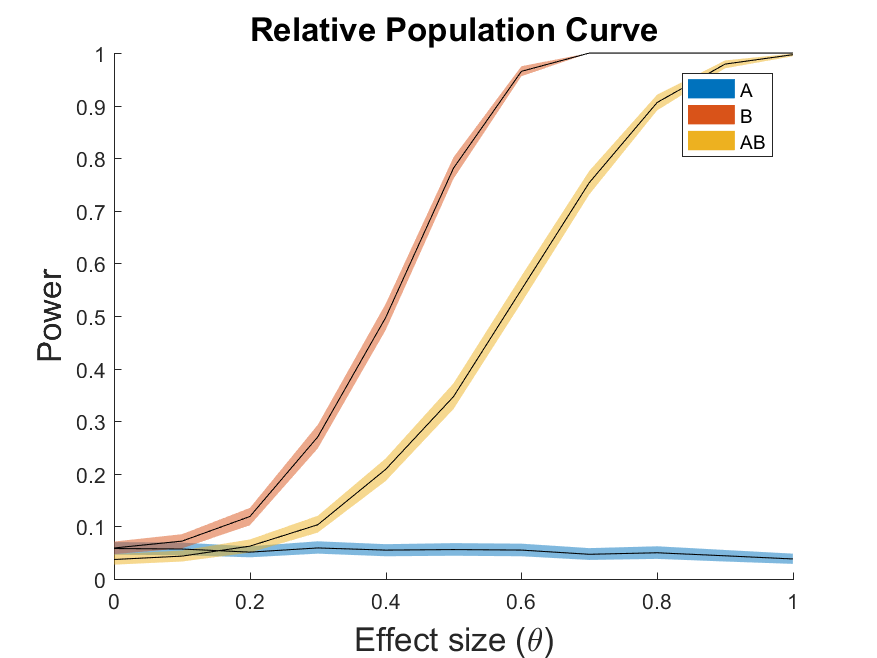}}
  	\caption{Examples of Relative Population Curve for model (\ref{eqn:RepMes}). The design matrix $\mathbf{F}$ contains a full factorial design with four levels in $A$, three levels in $B$ and four individuals in each cell of $C(A)$. Other inputs are $M=400$, $k_E = 1$, $R = 1000$, $P = 200$, $\delta=0.1$ and $\alpha = 0.05$. }

\label{fig:RPC2}
\end{figure*}

\subsection{Absolute Population Curves}

Figure \ref{fig:RPC3} shows four examples of APCs computed from the same parameters as the RPC in Figure \ref{fig:RPC1} and for $\theta=0.5$. The first example of APCs iteratively replicates the whole experiment (Figure \ref{fig:RPC3}(a)) and the remaining examples iteratively increase the number of levels of each of the factors (Figures \ref{fig:RPC3}(b), \ref{fig:RPC3}(c) and \ref{fig:RPC3}(d) for factors $A$, $B$ and $C(A)$, respectively). For reference, we marked with a red dashed line the same baseline situations in all APCs and the original RPC in Figure \ref{fig:RPC1}. Thus, the dashed line at $\theta=0.5$ in Figure \ref{fig:RPC1} identifies the same simulation point as at $\eta = 1$ in Figure \ref{fig:RPC3}(a), $\eta = 4$ (for four levels in $A$)
in Figure \ref{fig:RPC3}(b), $\eta = 3$ (for three levels in $B$)
in Figure \ref{fig:RPC3}(c), and $\eta = 4$ (for four replicates in $C(A)$, $r_{C{A}}=4$)
in Figure \ref{fig:RPC3}(d).

In general, we can see that increasing the replicates enhances the power in all factors and the interaction for all APCs. As discussed in Figure \ref{fig:xplan}, this enhancement is motivated by a reduction in the variance of the null and the alternative distributions of the test. Let us discuss this for each of the four examples.

\begin{figure*}
 	\centering
	\subfigure[Whole experiment replicated]{\includegraphics[width=0.45\textwidth]{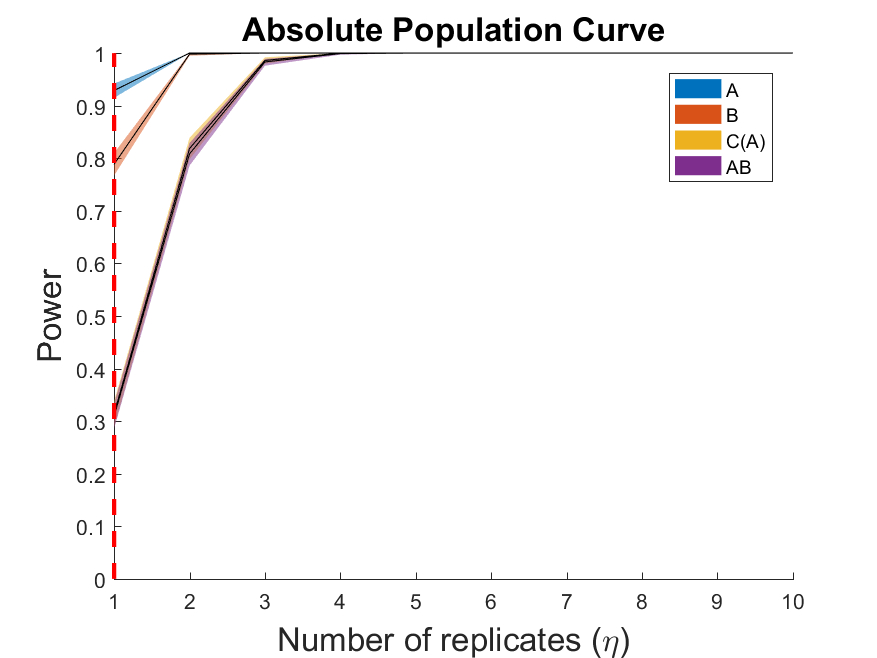}}
	\subfigure[Factor $A$ replicated]{\includegraphics[width=0.45\textwidth]{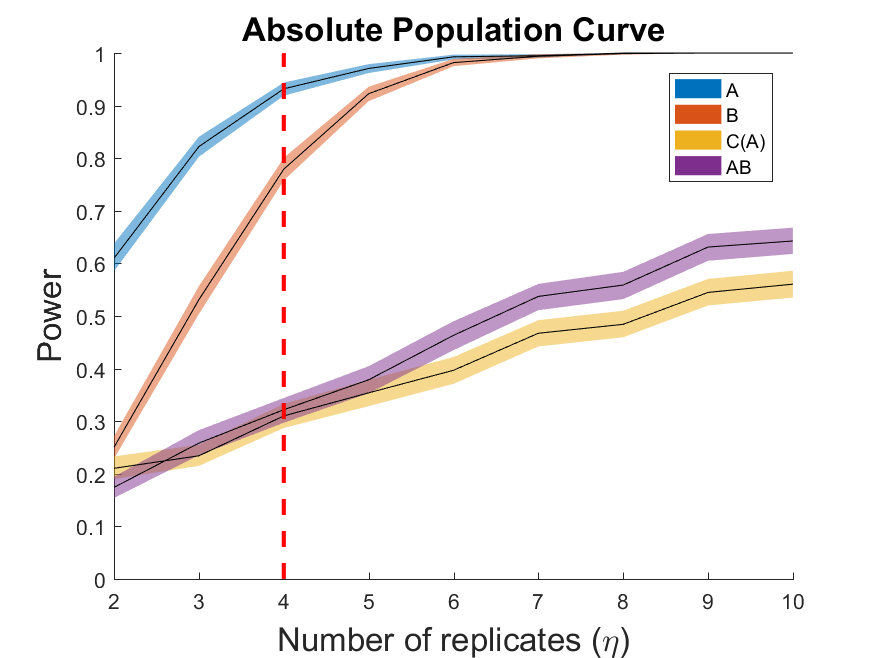}}
	\subfigure[Factor $B$ replicated]{\includegraphics[width=0.45\textwidth]{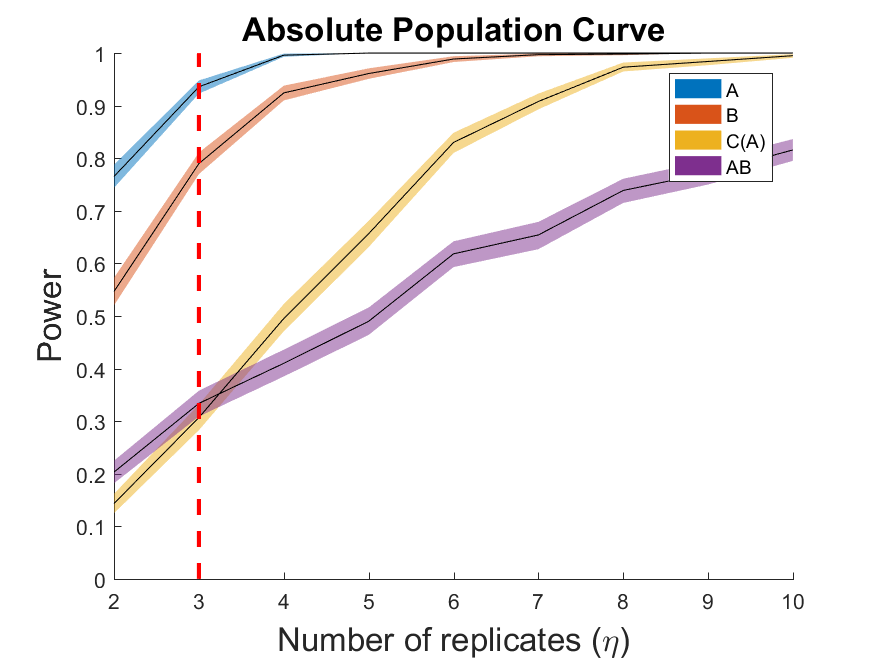}}
	\subfigure[Factor $C(A)$ replicated]{\includegraphics[width=0.45\textwidth]{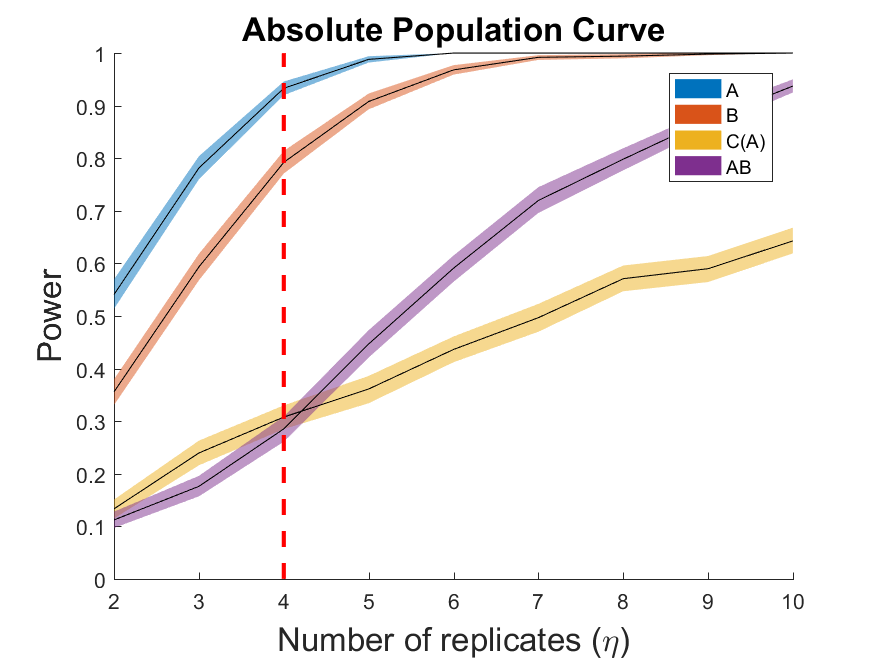}}
  {}	\caption{Examples of Absolute Population Curve for model (\ref{eqn:RepMes}). The original design matrix $\mathbf{F}$ contains a full factorial design with four levels in $A$, three levels in $B$ and four individuals in each cell of $C(A)$. Other inputs are $M=400$, $k_A = k_B = k_{C(A)} = k_{AB} = 0.2\theta$ for $\theta = 0.5$ and $k_E = 1$, $R = 1000$, $P = 200$ and $\alpha = 0.05$. }
\label{fig:RPC3}
\end{figure*}

In Figure \ref{fig:RPC3}(a), we duplicate the whole experiment, but all levels of the factors remain the same. This makes all variance coefficients in Eqs. {(\ref{eq:EA})-(\ref{eq:pool})}, with the exception of the variance of the error, to be multiplied by a factor of 2 (and in general of $\eta$ if the experiment is duplicated $\eta$-wise). This makes the expected MS's, and in turn the expected F-ratios, larger. The null distribution of a single instance simulated with this duplication (Figure \ref{fig:Null2}(a)) remains similar to the original one with no duplication in Figure \ref{fig:Null}. {However, if we compare the ASCA tables of the case with and without duplication, Tables \ref{table:4_1} and \ref{table:ASCA}, respectively, we can see that after duplication all the factors and the interaction are statistically significant as a result of the larger F-ratios. This is correctly depicted by the APC in Figure \ref{fig:RPC3}(a), where power for $\eta=2$ is above 0.8 for all factors and the interaction (which means that in more than 80\% of the simulated experiments we get statistically significant differences in the factors and the interaction).}   

\begin{figure*}
 	\centering
	\subfigure[]{\includegraphics[width=0.45\textwidth]{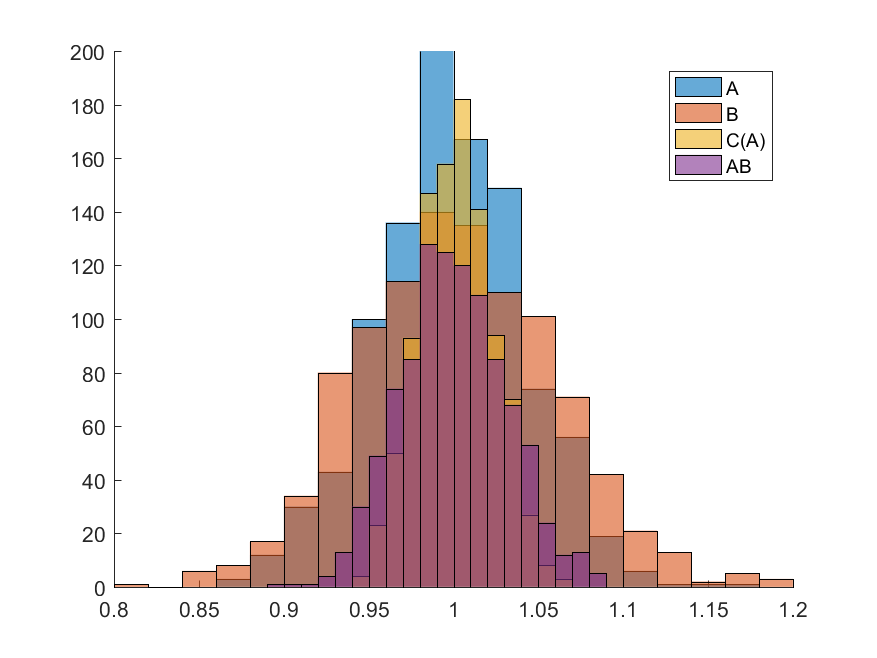}}
	\subfigure[]{\includegraphics[width=0.45\textwidth]{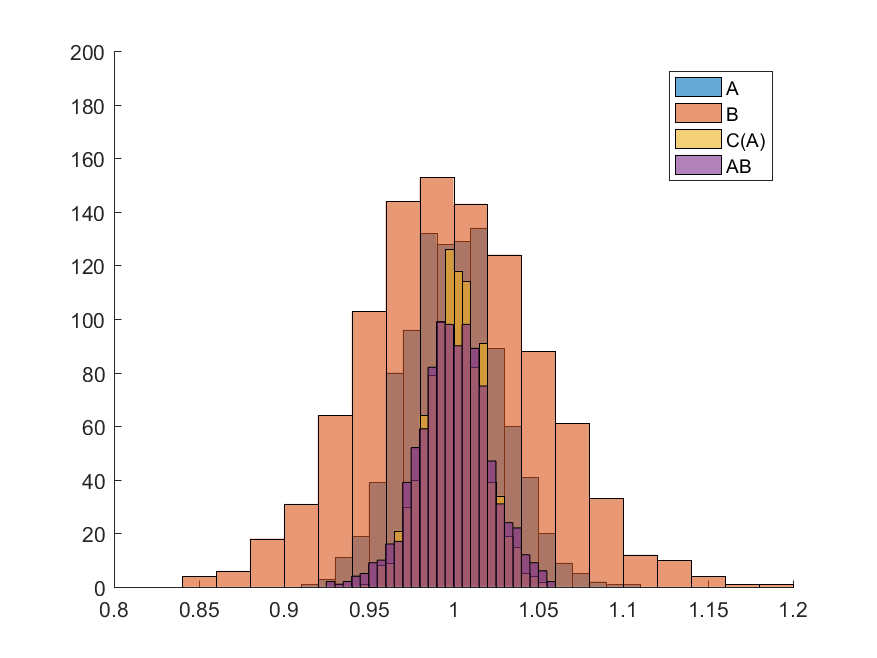}}
	\subfigure[]{\includegraphics[width=0.45\textwidth]{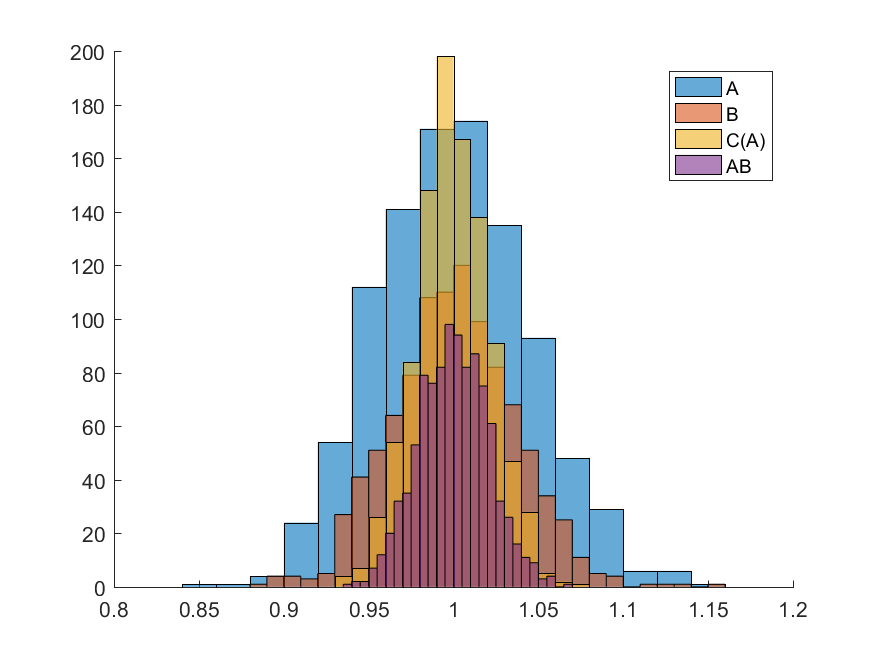}}
	\subfigure[]{\includegraphics[width=0.45\textwidth]{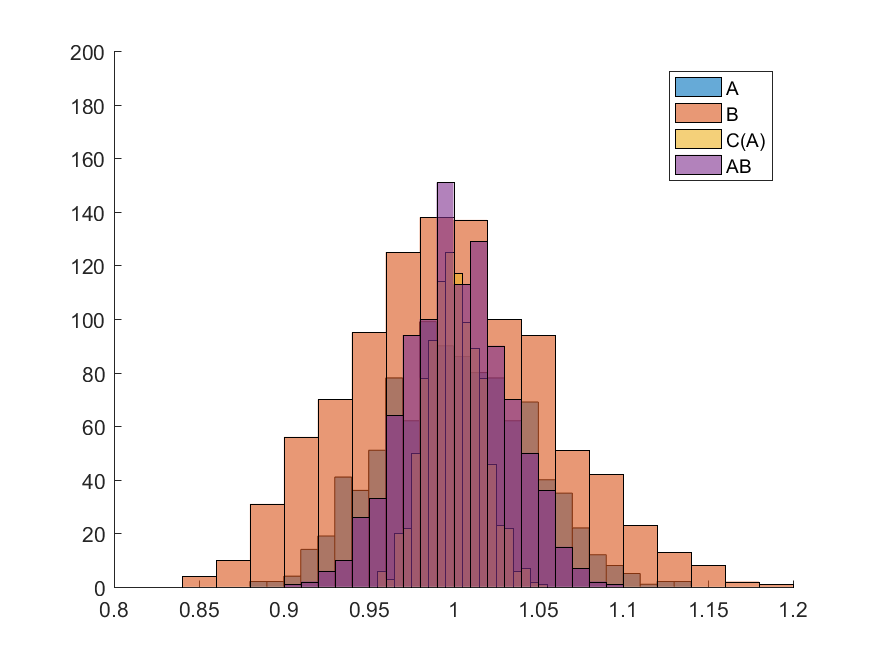}}

 	\caption{Null distributions for the first simulated dataset in the RPC in Figure \ref{fig:RPC1} and for $\theta = 0.5$, but when the Whole experiment replicated (a), the number of levels of $A$ is duplicated (b), the number of levels of $B$ is duplicated (c), and the number of levels of $C(A)$ is duplicated (d).  }
\label{fig:Null2}
\end{figure*}

In Figure \ref{fig:RPC3}(b), the APCs show the behaviour of the power in terms of the number of levels of $A$, $L_A$. From all of the expected MS values, only $E(MS_B)$ in Eq. (\ref{eq:B}) is affected by $L_A$. Consequently, the APC with larger slope in terms of $L_A$ is actually $B$. All the other factors and the interaction are increase their power with $L_A$, but rather than because of a change in expected MS and/or F-tatio, they do because of a change in their null distribution. This can be seen by comparing Figure \ref{fig:Null}, for $L_A = 4$, with Figure \ref{fig:Null2}(b), for $L_A = 8$. The latter shows a clear reduction in the variance of the null distributions of {$A$, $C(A)$ and $AB$. This reduction of variance leads to an increase of power.   
Finally, comparing Tables \ref{table:ASCA} and \ref{table:4_2}, we can see that the duplication of the levels of $A$ clearly reduces the p-value in $A$ and $B$, in agreement to what we see in in Figure \ref{fig:RPC3}(b), where the power of these factors for $\eta=8$ is close to 1 (that is, in almost 100\% of the simulated experiments we get statistically significant differences in these factors, but only 50\% for $C(A)$ and $AB$).}

Figures \ref{fig:RPC3}(c) and \ref{fig:RPC3}(d) illustrate that the power in factor $C(A)$ is mostly affected by the increase of levels of $B$, $L_B$, and the power of the interaction $AB$ is mostly affected by the replicates in $C(A)$, $r_{C(A)}$. All the other factors and interactions also increase their power. Again, this increase of power is a complex mixture of a modification of the expected MS's and F-ratios, and a reduction of the variance in the null distribution. { It is hopeless to predict this behaviour without computational means, but  easily observed  in Figures \ref{fig:Null2}(c) and \ref{fig:Null2}(d) and Tables \ref{table:4_3} and \ref{table:4_4}.} 

As a general conclusion, we can see that sampling size in the form of replicates and varying number of levels in the factors can have a complex influence on the relative power of factors and their interactions. In complex practical cases, an easy way to understanding how any form of duplication affect the power of each given factor and/or interaction is through the APC algorithm.

\section{Conclusion} \label{sec:Conclusion}

In this paper, we introduce the population power curves for ASCA and demonstrate them in simulation, discussing their relation to the theory of ANOVA and derive two useful forms of curve: Relative Population Curves (RPCs) and Absolute Population Curves (APCs). RPCs are useful to find the optimal ASCA pipeline for the analysis of an experimental design at hand. APCs are useful to determine the sample size and the optimal number of levels for each factor during the planning phase on an experiment. We believe that both tools should be adopted by ASCA practitioners to plan their experimental design (APCs) and analysis pipeline (RPCs) {during the planning phase of a multivariate experimental design.}

{In a sequel of this paper, we will introduce the sample power curves for ASCA, which is an optimized version of a power curve when a small sample of the experiment at hand is available, for instance obtained by running a reduced number of trials before a larger experiment.}

\section*{Software} 
All simulated examples in the paper can be reproduced with the code available at \url{https://github.com/CoDaSLab/PopulationCurvesASCA}

\section*{Acknowledgement}
\label{sec:Acknowledgments}
	
This work was supported by the Agencia Estatal de Investigación in Spain, MCIN/AEI/ 10.13039/501100011033, grant No PID2020-113462RB-I00.  Michael Sorochan Armstrong has received funding from the European Union’s Horizon
Europe research and innovation programme under the Marie Skłodowska- Curie grant agreement No 101106986. Funding for open access charge: Universidad de Granada / CBUA. We would like to acknowledge the feedback and interesting comments on the paper of Prof. Morten Rasmussen.

\appendix 

\section{Theoretical derivation of expected F-ratios} \label{Math}

Given that all factors are derived from a (pseudo-)random distribution, and following Montgomery~\cite{montgomery2020design}, we can derive the relationships between population variances and expected MS's in model (\ref{eqn:RepMes}):  
		\begin{equation}\label{eq:res}
		E(MS_{E}) = \sigma^2_E 
		\end{equation}
		\begin{equation}\label{eq:EA}
		E(MS_{AB}) = \sigma^2_E +  r_{C(A)} \cdot \sigma_{AB}^2 
		\end{equation}
		\begin{equation}
		E(MS_{C(A)}) = \sigma^2_E + L_B  \cdot \sigma^2_{C(A)} 
		\end{equation}
		\begin{equation}
		E(MS_{B}) = \sigma^2_E + r_{C(A)} \cdot \sigma_{AB}^2 + L_A \cdot r_{C(A)} \cdot \sigma_{B}^2
		\end{equation}
		\begin{equation}\label{eq:pool}
		E(MS_{A}) = \sigma^2_E + r_{C(A)} \cdot \sigma_{AB}^2 + L_B  \cdot \sigma^2_{C(A)} + L_B \cdot r_{C(A)}  \cdot \sigma_{A}^2 
		\end{equation}
where $\sigma^2_A$, $\sigma^2_B$, $\sigma^2_{C(A)}$, $\sigma^2_{AB}$, and $\sigma^2_E$ are the population variance of the factors, the interaction and the residuals, respectively, and in our example we have $r_{C(A)} = 4$ replicates, and number of levels $L_A = 4$ and $L_B = 3$.

Combining previous equations and Eqs. (\ref{eq:f}) to (\ref{eq:approx}), the inference statistics follow:
        \begin{dmath}
            \label{eq:approxA}
		E(F_A) = \frac{E(MS_{A})}{(DoF_{C(A)} E(MS_{C(A)}) + DoF_{AB} E(MS_{AB}))/(DoF_{C(A)} + DoF_{AB})}  =  \frac{(\sigma^2_E + r_{C(A)} \cdot \sigma_{AB}^2 + L_B  \cdot \sigma^2_{C(A)} + L_B \cdot r_{C(A)}  \cdot \sigma_{A}^2)(r_{C(A)} + L_B-1)}{r_{C(A)} (\sigma^2_E + L_B  \cdot \sigma^2_{C(A)}) + (L_B-1) (\sigma^2_E +  r_{C(A)} \cdot \sigma_{AB}^2)}
		\end{dmath}
		\begin{equation}
            \label{eq:B}
		E(F_B) = \frac{E(MS_{B})}{E(MS_{AB})} =  1 + \frac{L_A \cdot r_{C(A)} \cdot \sigma_{B}^2}{\sigma^2_E + r_{C(A)} \cdot \sigma_{AB}^2}
		\end{equation}
	    \begin{equation}
            \label{eq:C}
		E(F_C(A)) = \frac{E(MS_{C(A)})}{E(MS_{E})} = 1 + \frac{L_B  \cdot \sigma^2_{C(A)}}{\sigma^2_E}
		\end{equation}
		\begin{equation}
            \label{eq:AB}
		E(F_{AB}) = \frac{E(MS_{AB})}{E(MS_{E})} = 1 + \frac{r_{C(A)} \cdot \sigma_{AB}^2}{\sigma^2_E}
		\end{equation}
Adjusting the population variances to the square of the standard deviation coefficients used in the RPCs of Figure \ref{fig:RPC1}, i.e., {\color{blue} $\sigma^2_A = \sigma^2_B = \sigma^2_{C(A)} = \sigma^2_{AB} = 0.04\theta^2$ and $\sigma^2_E = 1$, it holds:
        \begin{equation}\label{eq:approxA2}
		E(F_A) =  \frac{6(1 + 0.76 \theta^2)}{4 (1 + 0.12 \theta^2) + 2 (1 +  0.16 \theta^2)}
		\end{equation}
		\begin{equation}
		E(F_B) =  1 + \frac{0.64 \theta^2}{1 + 0.16 \theta^2}
		\end{equation}
	    \begin{equation}
		E(F_C(A)) =  1 + 0.12  \theta^2
		\end{equation}
		\begin{equation}
            \label{eq:AB2}
		E(F_{AB}) = 1 + 0.16 \theta^2
		\end{equation}

From Eqs. (\ref{eq:B}), (\ref{eq:C}) and (\ref{eq:AB}), we can see that if we set $\sigma^2_B = 0$, $\sigma^2_{C(A)} = 0$ and/or $\sigma^2_{AB} = 0$, the corresponding expected F-ratio equals 1 regardless the variance of the error. This makes the RPC to adjust to the expected type I error regardless of $\theta$.} This behaviour is not found for factor A. The reason can also be found in the corresponding equation of the F-ratio, see Eq. (\ref{eq:approxA}). This equation represents an approximate test rather than an exact one \cite{anderson2003permutation, montgomery2020design}. If we set $\sigma^2_A = 0$, and adjust the remaining population variances to the square of the standard deviation coefficients in terms of $\theta$, it now holds:
{\color{blue}%
        \begin{equation}
		E(F_A) =  \frac{6(1 + 0.24 \theta^2)}{4 (a + 0.12 \theta^2) + 2 (1 +  0.16 \theta^2)}
		\end{equation}
  }

  \section{Tables} \label{App}

ASCA tables for a single instance (dataset) simulate with the same parameters of the RPC in Figure \ref{fig:RPC1} and for $\theta = 0.5$, but i) when the whole experiment is duplicated (Table \ref{table:4_1}); ii) when the number of levels of $A$, $L_A$, is duplicated  (Table \ref{table:4_2}); iii) when the number of levels of $AB$, $L_B$, is duplicated  (Table \ref{table:4_3}); and iv) when the number of replicates in $C(A)$, $r_{C(A)}$, is duplicated  (Table \ref{table:4_4}); 

{\color{blue}
\begin{table} 
\begin{tabular}{llllllll}
 & SumSq & PercSumSq & df & MeanSq & F & Pvalue \\ 
 \hline 
Mean & 1.7243 & 1.7257 & 1 & 1.7243 &  &  \\ 
A & 3.8883 & 3.8914 & 3 & 1.2961 & 1.1885 & 0.000999 \\ 
B & 2.7994 & 2.8017 & 2 & 1.3997 & 1.2888 & 0.000999 \\ 
C(A) & 13.1133 & 13.1239 & 12 & 1.0928 & 1.0946 & 0.000999 \\ 
AB & 6.5162 & 6.5215 & 6 & 1.086 & 1.0879 & 0.001998 \\ 
Residuals & 71.8774 & 71.9358 & 72 & 0.9983 &  &  \\ 
Total & 99.9188 & 100 & 96 & 1.0408 &  &  \\ 
\end{tabular} 
\caption{ASCA table for the first simulated dataset in the RPC of Figure \ref{fig:RPC1} and for $\theta = 0.5$, when the whole experiment is duplicated.}
\label{table:4_1}
\end{table} 

\begin{table} 
\begin{tabular}{llllllll}
 & SumSq & PercSumSq & df & MeanSq & F & Pvalue \\ 
 \hline 
Mean & 1.6805 & 1.6809 & 1 & 1.6805 &  &  \\ 
A & 8.1368 & 8.1388 & 7 & 1.1624 & 1.1314 & 0.000999 \\ 
B & 2.6052 & 2.6058 & 2 & 1.3026 & 1.2794 & 0.000999 \\ 
C(A) & 24.7872 & 24.7931 & 24 & 1.0328 & 1.0219 & 0.1049 \\ 
AB & 14.2541 & 14.2575 & 14 & 1.0182 & 1.0074 & 0.37363 \\ 
Residuals & 48.5123 & 48.5239 & 48 & 1.0107 &  &  \\ 
Total & 99.9762 & 100 & 96 & 1.0414 &  &  \\ 
\end{tabular} 
\caption{ASCA table for the first simulated dataset in the RPC of Figure \ref{fig:RPC1} and for $\theta = 0.5$, when the number of levels of $A$ is duplicated.}
\label{table:4_2}
\end{table} 

\begin{table} 
\begin{tabular}{llllllll}
 & SumSq & PercSumSq & df & MeanSq & F & Pvalue \\ 
 \hline 
Mean & 1.537 & 1.5426 & 1 & 1.537 &  &  \\ 
A & 3.9359 & 3.9501 & 3 & 1.312 & 1.2495 & 0.000999 \\ 
B & 5.8933 & 5.9146 & 5 & 1.1787 & 1.1578 & 0.000999 \\ 
C(A) & 13.0792 & 13.1262 & 12 & 1.0899 & 1.0913 & 0.000999 \\ 
AB & 15.2703 & 15.3253 & 15 & 1.018 & 1.0193 & 0.17283 \\ 
Residuals & 59.9256 & 60.1413 & 60 & 0.99876 &  &  \\ 
Total & 99.6414 & 100 & 96 & 1.0379 &  &  \\ 
\end{tabular} 
\caption{ASCA table for the first simulated dataset in the RPC of Figure \ref{fig:RPC1} and for $\theta = 0.5$, when the number of levels of $B$ is duplicated.}
\label{table:4_3}
\end{table} 

\begin{table} 
\begin{tabular}{llllllll}
 & SumSq & PercSumSq & df & MeanSq & F & Pvalue \\ 
 \hline 
Mean & 1.6665 & 1.6705 & 1 & 1.6665 &  &  \\ 
A & 4.0614 & 4.0712 & 3 & 1.3538 & 1.3083 & 0.000999 \\ 
B & 2.6655 & 2.672 & 2 & 1.3328 & 1.2989 & 0.000999 \\ 
C(A) & 29.0265 & 29.0969 & 28 & 1.0367 & 1.0333 & 0.021978 \\ 
AB & 6.1566 & 6.1716 & 6 & 1.0261 & 1.0228 & 0.24276 \\ 
Residuals & 56.1816 & 56.3178 & 56 & 1.0032 &  &  \\ 
Total & 99.758 & 100 & 96 & 1.0391 &  &  \\ 
\end{tabular} 
\caption{ASCA table for the first simulated dataset in the RPC of Figure \ref{fig:RPC1} and for $\theta = 0.5$, when the number of replicates in $C(A)$ is duplicated.}
\label{table:4_4}
\end{table} 

}

\afterpage{\clearpage}
\bibliography{Bibliography}
\bibliographystyle{ieeetr}

\end{document}